\documentclass{ws-ijmpa}
\begin{document}
\def\ds{\displaystyle}
\def\f{\varphi}
\def\d{\delta}
\def\p{\partial}
\markboth{M. B.~Golubev \& S. R.~Kelner}{Point Charge Self-Energy
in the General Relativity} \catchline{}{}{}{}{}
\title{POINT CHARGE SELF-ENERGY IN THE GENERAL RELATIVITY}
\author{\footnotesize M. B. GOLUBEV}
\address{Russian Federal Nuclear Center VNIIEF\\
Mira, 37, Sarov, 607200  Nizhegorodskaya, Russia
\\ muhomor@vniief.ru}
\author{\footnotesize S. R. KELNER}
\address{Moscow Engineering Physics Institute\\
Kashirskoe, 31, 115409 Moscow, Russia
\\ mike\_k@hotbox.ru}
\maketitle
\pub{Received 22 October 2004}{}
\begin{abstract}
Singularities in the metric of the classical solutions to the
Einstein equations (Schwarzschild, Kerr, Reissner -- Nordstr\"om
and Kerr -- Newman solutions) lead to appearance of generalized
functions in the Einstein tensor that are not usually taken into
consideration. The generalized functions can be of a more complex
nature than the Dirac $\d$-function. To study them, a technique
has been used based on a limiting solution sequence. The solutions
are shown to satisfy the Einstein equations everywhere, if the
energy-momentum tensor has a relevant singular addition of
non-electromagnetic origin. When the addition is included, the
total energy proves finite and equal to $mc^2$, while for the Kerr
and Kerr--Newman solutions the angular momentum is $mc\,{\bf a}$.
As the Reissner--Nordstr\"om and Kerr--Newman solutions correspond
to the point charge in the classical electrodynamics, the result
obtained allows us to view the point charge self-energy divergence
problem in a new fashion.
\end{abstract}
\keywords{Self-energy; classical electron; point charge.}
\section*{Introduction}\label{intr}
A major general relativity principle is the equality of inert and
gravitating masses. However, the classical solutions to the
Einstein equations (Schwarzschild, Kerr, Reissner--Nordstr\"om and
Kerr-Newman solutions) do not satisfy that principle at first
sight. For the Schwarzschild and Kerr solutions the
energy-momentum tensor and, hence, self- energy are zero, for the
Reissner--Nordstr\"om and Kerr--Newman solutions the self-energy
is infinite, whereas the gravitating mass is finite for all these
solutions. A reason for this unconformity can be that the above
solutions satisfy the Einstein equations not in the entire space.
A property common to all of the solutions is existence of
singularity of form \mbox{$\sim1/r$}, \mbox{$\sim1/r^2$} in the
metric. This suggests that the Einstein tensor containing second
derivatives of the metric can contain the generalized functions,
which are lost in the direct differentiation and, therefore, are
not included in the energy-momentum
tensor.\footnote{\parbox[t]{5in} {For example, in electrostatics
for the point charge potential we have $\triangle
\f=-4\,\pi\,e\,\d({\bf r})$, while\\the direct differentiation
yields $\ds\frac1{r^2}\frac{\p}{\p r}\left(r^2\,\frac{\p\f}{\p
r}\right)=0$.}} The paper demonstrates that the Einstein tensor
for the aforementioned solutions in fact contains the generalized
functions, which can be of a more complex nature than the Dirac
$\d$-function. If we require validity of the Einstein equations in
the \textit{entire} space, including $r=0$, then an appropriate
singular term must be added to the
energy-momentum tensor.\\[-24pt]

\section{The Analogy with Electrostatics}\label{El}
It is simplest to elucidate the method determining if the
generalized function appears in the singular function
differentiation by the example of
electrostatics. The point charge potential\\[-16pt]

\begin{equation}\label{phi}
    \f=\frac er
\end{equation}
is singular at point $r=0$ and satisfies Poisson equation:

\begin{equation}
\label{Puas1}
\triangle\f=-4\,\pi\rho\,,
\end{equation}
where $\rho=e\,\d({\bf r})$. A method to ascertain this is the
following. Replace potential (\ref{phi}) by the nonsingular
function of form
\begin{equation}\label{PC}
\tilde \f= \frac{e}{r}\,\,\theta(r-r_0)+
\left(\frac{3\,e}{2\,r_0}-\frac{e\,r^2}{2\,r_0^3}\right)\theta(r_0-r)\,,
\end{equation}
where $\theta(x)$ is Heaviside function ($\theta(x)=1$ for $x>0$
and $\theta(x)=0$ for $x<0$). Having substituted this potential
into (\ref{Puas1}), we find that $\tilde\f$ will be a solution to
the Poisson equation for charge density
\begin{equation}\label{rho}
\rho=\frac{3\,e}{4\pi\,r_0^3}\,\,\theta(r_0-r)\,.
\end{equation}
The integral of (\ref{rho}) over volume is independent of $r_0$
and equal to $e$. In the limit $r_0\to0 $, $\tilde\f\to e/r$ ,
$\rho\to e\,\d({\bf r})$, i.e. the limit of solution (\ref{PC})
corresponds to the presence of a point source with charge $e$ in
the origin of coordinates and is a solution to equation
(\ref{Puas1}). It is easy to show that this result is independent
of the choice of the potential in range $r<r_0$, with the smooth
behavior of the potential at point $r=r_0$ being not necessary.
The result is always single: in the limit $r_0\to0 $, the
potential is $\f=\ds\frac{e}{r}$ and the charge density is $\rho=
e\,\d({\bf r})$. Below we apply a similar procedure to the
classical solutions of the Einstein equations.

\section{Self-Energy}
\label{SE}

What should be meant by the self-energy in the general relativity
is not a trivial question. This question is typically solved using
the energy-momentum pseudotensor (see, e.g., in
Ref.~\refcite{Xulu} and references thereof). A demerit of the
approach is that the system self-energy definition is related to a
special (Cartesian) system of coordinates and is not invariant
under the coordinate transformations. The energy-momentum
pseudotensor allows energy density to be assigned to the
gravitational field; the energy density, however, cannot be
localized.

A self-energy definition can be suggested based on the
energy-momentum tensor of fields and material only. For stationary
and static solutions there is Killing vector $\xi_0=\p/\p t$
generating conserved current $J^\mu=T^\mu_\nu\,\xi_0^\nu$, where
$\xi_0^\nu=(1,{\bf 0})$ are the contravariant vector
components\cite{Logunov}. As $\nabla_\mu\,J^\mu=0$, conservation
law
\begin{equation}
\frac{d}{dt}\int\!d^3x\,\sqrt{-g}\,J^0=\int\!
dS_k\,\sqrt{-g}\,J_0^k\,.
\end{equation}

is satisfied. If the energy density is defined as a zero component
of the current, then total energy

\begin{equation}
 E=\int\! d^3x\,\sqrt{-g}\,J^0=\int\! d^3x\,\sqrt{-g}\,T_0^0
\end{equation}

will be independent of a choice of the system of coordinates.

\section{Reissner--Nordstr\"om Metric}\label{MRN}

The Reissner--Nordstr\"om solution is of the form\cite{Misner}
\begin{equation}\label{RN}
ds^2=\frac{\Phi}{r^2}\,dt^2
 -\frac{r^2\,dr^2}{\Phi}
 -r^2\left(\sin^2\vartheta\,d\f^2+d\vartheta^2\right)\,,
\end{equation}

where $\Phi=r^2-2\,m\,r+Q^2$ ($m$ and $Q$ are the mass and charge,
respectively\footnote{%
Those units are used, in which the gravitational constant and
light speed are equal to 1.}). This solution satisfies Einstein
equations
\begin{equation}
\label{Ein}
 G^{\mu\nu}=8\pi\,T^{\mu\nu}\,,
\end{equation}

where $\ds T^{\mu\nu}=\frac1{4\pi}\big({F^\mu}_\alpha
F^{\alpha\nu}+\frac14\,g^{\mu\nu}
F_{\alpha\beta}F^{\alpha\beta}\big)$ is the electromagnetic field
energy- momentum tensor everywhere, except for point $r=0$, at
which the solution is singular. The singularity structure of the
tensor $ G^{\mu\nu}$ and nature of the appearing generalized
function can be found out using a procedure similar to that
described in Section \ref{El}.

Consider the metric of form (\ref{RN}), having substituted the
following function for $\Phi$ in it:
\begin{equation}\label{SMod}
\widetilde\Phi= \left(r^2-2\,m\,r+Q^2\right)\,\theta(r-r_0)+
\frac{r^2}{r_0^2}\left( r_0^2-2\,m\,
r_0\,+Q^2\right)\,\theta(r_0-r)\,.
\end{equation}

In so doing the metric becomes non-singular and in the limit
$r_0\to0$ transfers to metric (\ref{RN}). The energy-momentum
tensor corresponding to the metric can be derived from the
Einstein equations. The (0,0) component of the tensor is

\begin{equation}\label{Tmnmod}
 T^0_0=\frac1{8\,\pi}\,G^0_0=
\frac{Q^2}{8\pi\,r^4}\,\theta(r-r_0)+
\left(-\frac{Q^2}{8\pi\,r^2\,r_0^2}+\frac
m{4\pi\,r^2\,r_0}\right)\theta(r_0- r)\,.
\end{equation}

In this expression the first term is the electrostatic field
energy confined in range $r>r_0$. The second term appearing from
the metric smoothing does not disappear in the limit $r_0\to0$.
The self-energy in the solution constructed is
\begin{equation}
\label{ERN}
 E=\int\! d^3x\,\sqrt{-g}\,T_0^0=\frac{Q^2}{2\,r_0}+
 \left(-\frac{Q^2}{2\,r_0} +m \right)=m\,.
\end{equation}

Result (\ref{ERN}) can be shown to be independent of the metric
smoothing method. In the limit $r_0\to0$ relation (\ref{Tmnmod})
can be written as
\begin{equation}\label{Tmnmod1}
 T^0_0=\frac1{\sqrt{-g}}\left( m\,\d({\bf r})+\frac12\,Q^2\varpi({\bf
r})\right).
\end{equation}

Here $\varpi({\bf r})$ is the generalized function determined by
the following integration rule:
\[
\int\!f({\bf r})\,\varpi({\bf r})\,d^3x=\int\!\frac{f({\bf r})-
f(0)}{r^4}\,d^3x\,,
\]
where $f({\bf r})$ is a bounded smooth function.

For the Schwarzschild metric ($Q=0$ in (\ref{RN})) the term
$m\,\d({\bf r})$ in $T^0_0$ that corresponds to a point source can
be obtained straightforwardly when the presence of term
$\sim\triangle(1/r)$ in $G^0_0$ is considered. A more complicated
generalized function $\varpi({\bf r})$ appears as a source when
$Q\ne0$. It owes its origin to the presence of term
$\sim\triangle(1/r^2)$ in $G^0_0$. Thus, the Schwarzschild and
Reissner--Nordstr\"om solutions can be extended to the entire
space, if the point source is added to the energy-momentum tensor.

\section{Kerr-Newman Metric}

Apply the same procedure to the Kerr--Newman metric\cite{DKS}
\begin{equation}\label{KN}
ds^2=\eta_{\mu\nu}\,dx^\mu\,dx^\nu+\Psi\left(dt- \frac{(r^2 x_k+r
({\bf x \times  a})_k +a_k \,{\bf a x})\,dx^k}{r( r^2+a^2)}
 \right)^2.
\end{equation}

Here
\begin{equation}\label{Psi}
 \Psi=\ds\frac{-2\,m\, r
+Q^2}{r^2+\ds\frac{{\bf a x}^2}{r^2}}\,,
\end{equation}

where ${\bf a}$ is a space vector, $a$ is its module, $r$ is
determined by equation

\begin{equation}
r^4-r^2({\bf x}^2-a^2)-({\bf a x})^2=0\,.
\end{equation}

\begin{figure}[t]
\centerline{ \psfig{file=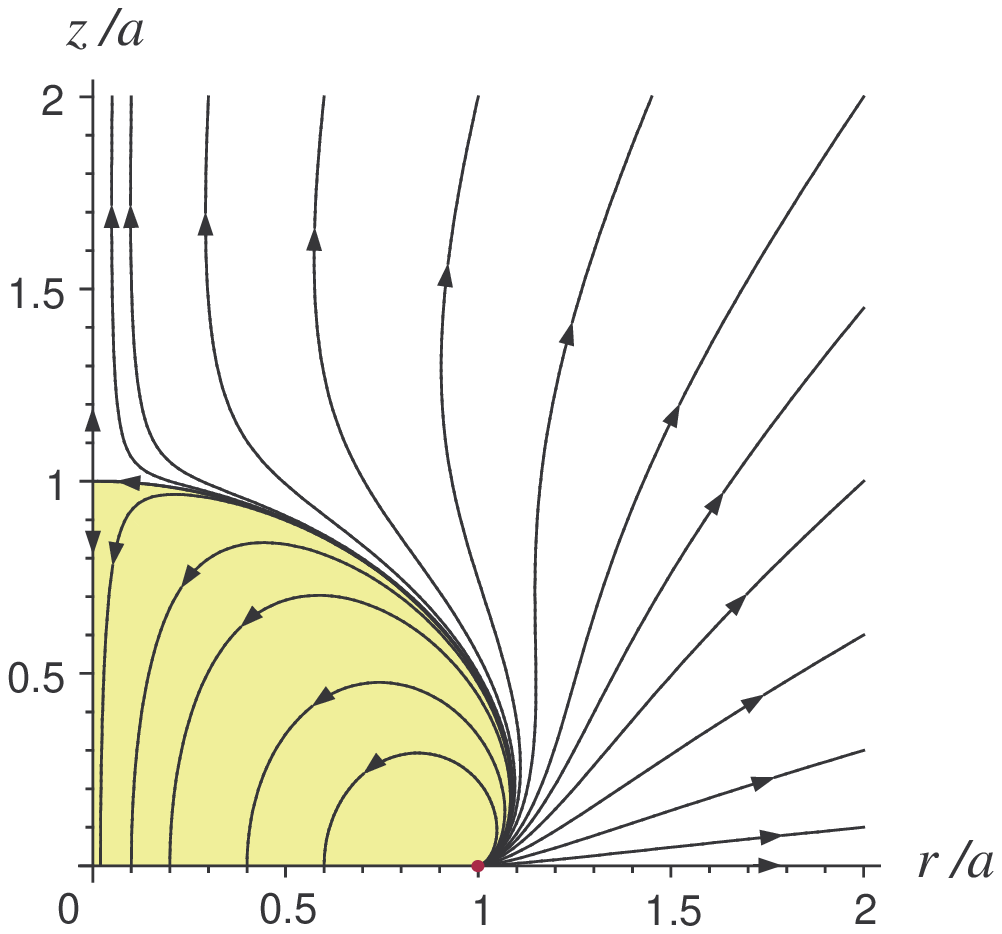,height=5.cm,clip=}
\hspace*{1cm} \psfig{file=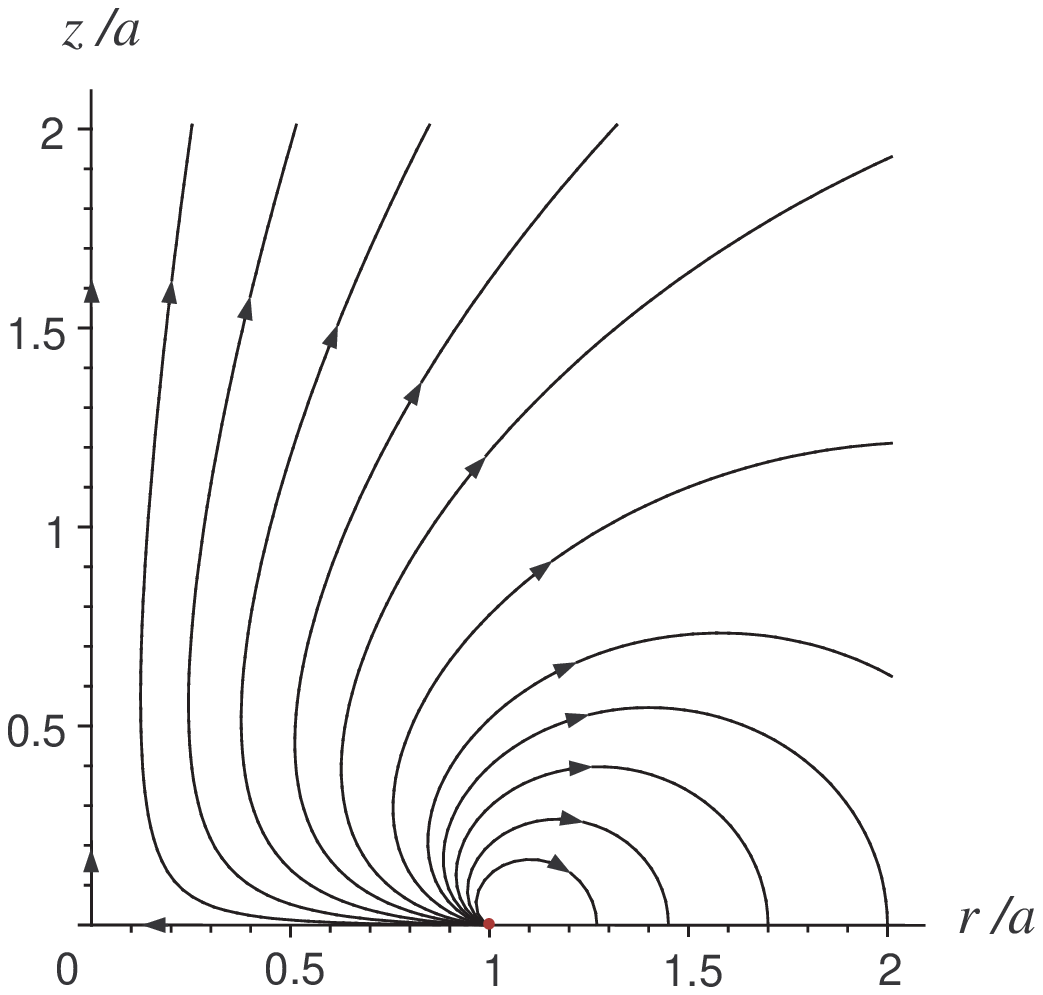,height=5.cm,clip=} }
\caption{Electric and magnetic field lines for the Kerr-Newman
solution.}\label{fig}
\end{figure}

It is easy to show that this is an ordinary Kerr-Schild
representation, if vector $\bf a$ is directed along axis $z$. The
surfaces of constant $r$ are ellipsoids of revolution, whose axis
coincides with the direction of the vector $\bf a$. With $r=0$ the
ellipsoid degenerates to a disk of radius $a$. The metric is
continuous on the disk, but the components of the metric and
electromagnetic field 4-potential and strength undergo a kink,
while the electromagnetic field strength a discontinuity (see
Fig.~\ref{fig}). This means that on the disk there is a singular
distribution of mass, charge, and currents, which is not embodied
in the electromagnetic field energy-momentum tensor.

Construct a solution similar to solution (\ref{SMod}). To do this,
in (\ref{KN}) replace $\Psi$ by function

\begin{equation}\label{KNMod}
\widetilde\Psi=\frac{-2\,m\, r+Q^2}{r^2+\ds\frac{{\bf a
x}^2}{r^2}}\,\theta(r-r_0)+ \frac{-2\,m\,
r_0+Q^2}{r_0^2+\ds\frac{{\bf a x}^2}{r_0^2}}\,\theta(r_0-r)\,.
\end{equation}

The constructed solution is continuous everywhere, but has a
derivative discontinuity at $r=r_0$. To satisfy the Einstein
equations, the (0,0)-component of the energy-momentum tensor
should be of the following form,
\[
T^0_0=\frac1{8\,\pi}\,G^0_0=
\frac{Q^2}{4\,\pi}\left(\frac{{a}^2}{{\rho }^3}-\frac{r^2}{{\rho
}^3}
   -\frac{1}{2\,{\rho }^2}\right)\,\theta(r-r_0)+
\]
\[
{r_0}^2\,\left(2\,M\,r_0-Q^2 \right) \,\frac{
    \left( {r_0}^4 - 3\,{\bf a  x}^2 \right) \,
    \left( r^4\,{r_0}^4 - {\bf a x}^4 +
      r^6\,a^2 + r^2\,{\bf a x}^2\,{a }^2 \right) }
    {8\,\pi\,r^2\,\left( r^4 + {\bf a x}^2 \right) \,
    {\left( {r_0}^4 + {\bf a x}^2 \right) }^3}\,\theta(r_0-r)+
\]

\begin{equation}\label{TKNmod}
\delta(r-r_0)\frac{r_0\,\left( M\,r_0\,
        \left( {r_0}^4 - 3\,{\bf a x}^2 \right)  +
        Q^2\,\left( -{r_0}^4 + {\bf a x}^2 \right)
        \right) \,\left( -{\bf a x}^2 +
        {r_0}^2\,{a}^2 \right)   }{8\,\pi\,
    {\left( {r_0}^4 + {\bf a x}^2 \right) }^3}\,,
\end{equation}

where $\rho=r^2+{\bf a x}^2/r^2$. The contribution to the
self-energy made by each of three parts of $T_0^0$ with $r_0\to0$
constitutes diverging quantities, but in the aggregate the
divergencies  are surprisingly compensated:
\[
E=\frac{Q^2}{4\,r_0} + \frac{Q^2\,\left( {r_0}^2 + {a }^2 \right)
\,
     \arctan (\frac{a }{r_0})}{4\,{r_0}^2\,a}+
\]
\[
\left(2\,m\,r_0 -Q^2 \right) \, \left(\frac{-3}{4\,r_0} +
    \frac{\left( 5\,{r_0}^2 + {a }^2 \right) \,
       \arctan (\frac{a }{r_0})}{4\,{r_0}^2\,a } \right)-
\]

\begin{equation}\label{last}
\frac{\left( 2\,Q^2 - 5\,m\,r_0 \right)}{2\,r_0}+ \frac{\left(
2\,Q^2\,r_0 -m( 5\,\,{r_0}^2 +
 {a }^2) \right) \,\arctan (\frac{a }{r_0})}
 {2\,r_0\,a }=\,m\,.
\end{equation}

The first, second and third lines in this relation are the
contributions of ranges $r>r_0$, $r<r_0$ and surface $r=r_0$,
respectively. It can be shown that, like for the
Reissner-Nordstr\"om metric, the result is independent of the
smoothing method. The result obtained also applies to the Kerr
metric (it will suffice to set $Q=0$ in (\ref{KN})).

The energy-momentum tensor also allows us to find the total moment
of the system. The contributions to the total moment made by
ranges $r>r_0$, $r<r_0$ and surface $r=r_0$ diverge in the limit
$r_0\to0$, but the divergencies are compensated and the total
moment proves equal to $mc\,{\bf a}$.

The difficulties that appeared in the derivation of the relations
in this paper have been overcome thanks to code Mathematica~5,
Wolfram Research, Inc.

\section{Conclusion}

As shown in this paper, for the Schwarzschild, Kerr,
Reissner--Nordstr\"om and Kerr--Newman solutions to satisfy the
Einstein equations in the entire space, including $r=0$, singular
terms containing generalized functions should be added to the
energy-momentum tensor. In so doing the total energy proves finite
and equal to $m\,c^2$ for any solution. For the nonzero-charge
solutions the addition plays the role of Poincare tensions, i.e.
infinite negative mass is located at the center.

The existence of negative mass for the nonzero-charge solutions is
also evidenced by the fact that for trial particles the
gravitational attraction transfers to repulsion even at classical
radius $\frac{Q^2}{m\,c^2}$. One can determine this having
analyzed the equations of motion of trial particles.

The presence of the singular terms in the energy-momentum tensor
follows from the formal requirement of the solution validity in
the entire space. The physical interpretation of the complete
energy-momentum tensor may require
involvement of other physical fields or matter.\\[-24pt]

\section*{Acknowledgements}

The work was carried out under the financial support by the
International Science and Technology Center (ISTC), Grant 1655.\\[-24pt]

\end{document}